\documentclass[twocolumn,secnumarabic,amssymb, nobibnotes, aps, prd]{revtex4-1}
\usepackage{amsmath}
\usepackage[utf8]{inputenc}
\usepackage{graphicx}
\usepackage{hyperref}
\RequirePackage[pagewise,mathlines]{lineno}

\newcommand{\der}{\mathrm{d}}
\newcommand{\rt}{{\mathbf{r}_T}}
\newcommand{\bt}{{\mathbf{b}_T}}
\newcommand{\kt}{{\mathbf{k}_T}}
\newcommand{\pt}{{\mathbf{P}_T}}

\usepackage{xcolor}

\begin{document}
\title{Imaging the nucleus with high-energy photons}%

\author{Spencer R. Klein}%
\email[]{srklein@lbl.gov}
\affiliation{Nuclear Science Division, Lawrence Berkeley National Laboratory, Berkeley, CA 94720 USA}

\author{Heikki M\"antysaari}%
\email[]{heikki.mantysaari@jyu.fi}
\affiliation{Department of Physics, University of Jyv\"askyl\"a P.O. Box 35, 40014 University of Jyv\"askyl\"a, Finland}
\affiliation{Helsinki Institute of Physics, P.O. Box 64, 00014 University of Helsinki, Finland}

\date{\today}%
\begin{abstract}
In the 1930's, nuclear physicists developed the first realistic atomic models, showing that nuclei were made up of protons and neutrons.  In the 1960's, Deep Inelastic Scattering experiments showed that protons and neutrons had internal structure: quarks and gluons (collectively, partons), and later experiments showed that the parton momentum distributions are different in heavy nuclei, compared to those in free nucleons.  This difference is not surprising; partons are sensitive to their environment, and two gluons from different nucleons may fuse together, for example.  

Understanding how quarks and gluons behave in the nuclear environment is a significant focus of modern nuclear physics.  Recent measurements have provided us with an improved understanding of how  quark and gluon densities are altered in heavy nuclei.   We have also begun to make multi-dimensional pictures of the nucleus, exploring how these alterations are distributed within heavy nuclei.  We naturally expect these modifications to be largest in the core of a nucleus, and smaller near its periphery; this can change the effective shape of the nucleus.  We have also started to explore the transverse momentum distribution of the partons in the nuclei, and, using incoherent photoproduction as a probe, study event-by-event fluctuations in nucleon and nuclei parton densities.  

This article will explore recent progress in  measurements of nuclear structure at high energy, with some emphasis on these multi-dimensional pictures.  We will also discuss how a future  electron-ion collider (EIC) with high luminosity and center-of-mass energy will make exquisitely detailed images of partons in a nucleus. 

\end{abstract}
\maketitle

\section{Introduction}

In a simple picture, atomic nuclei are made up of identical protons and neutrons, which are themselves made up of quarks and gluons.   Each nucleon comprises three up or down valence quarks, plus virtual gluons and sea quarks and antiquarks.  The gluons and sea quarks and antiquarks are ephemeral, flickering in and out of existence  \cite{Ellis:1991qj}.  The quarks and gluons are defined in terms of parton (quark or gluon) distribution functions, $q(x,Q^2)$ and $g(x,Q^2)$, where $x$ is the Bjorken momentum fraction - the fraction of the nucleon momentum carried by that parton, measured in the infinite momentum frame - and $Q^2$ is the momentum scale at which the measurement is made.  The larger the $Q^2$, the more detailed the picture. For example, a gluon observed at one $Q^2$ might have fluctuated into a quark-antiquark pair when viewed at higher $Q^2$.  Thus, the quark and gluon distributions evolve with changing $Q^2$. 

\begin{figure}[t]
\begin{center}
\includegraphics[width=0.5\textwidth]{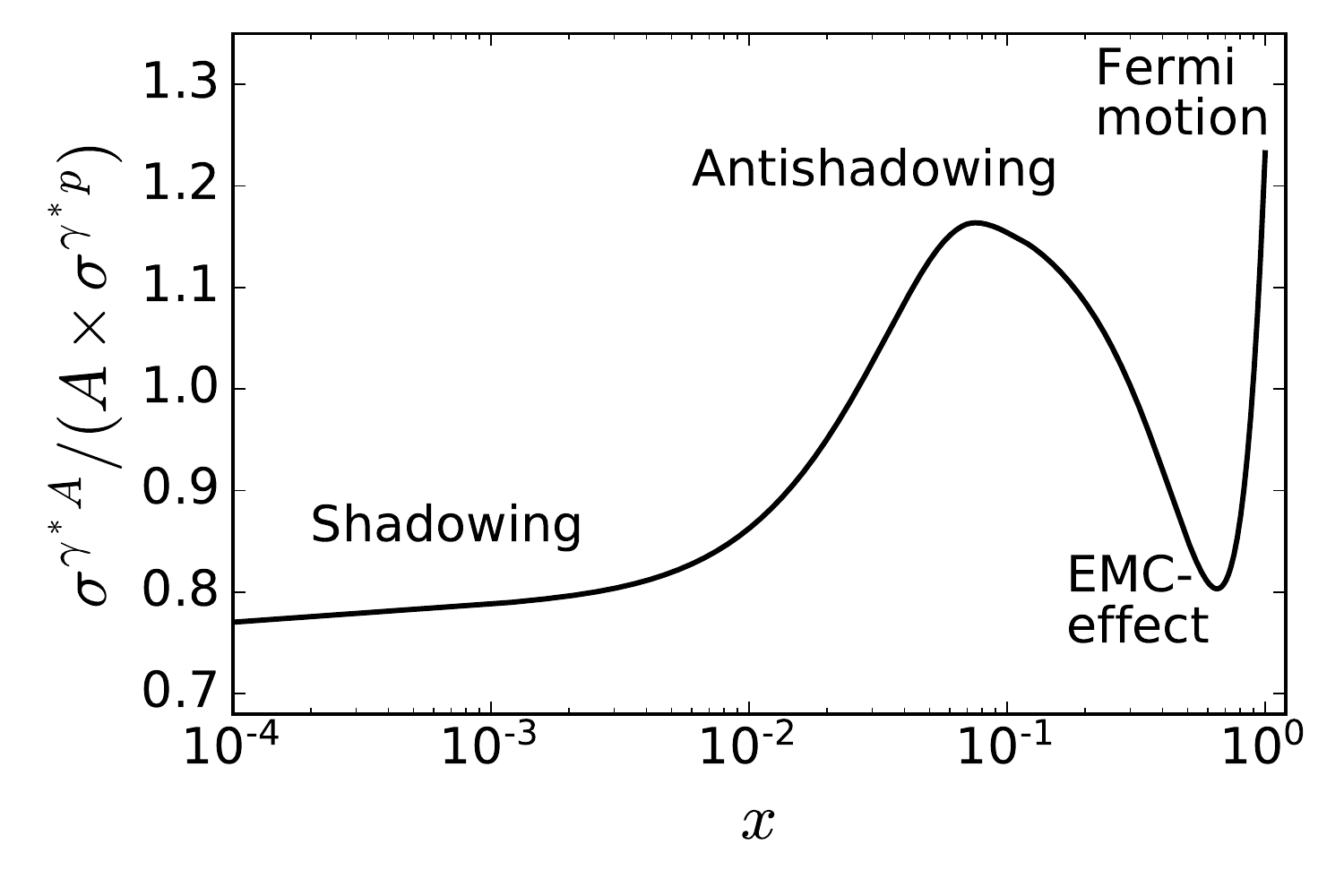}
\end{center}
\caption{Typical nuclear effects seen in deep inelastic scattering with nuclear targets. The cross section ratio  measures the nuclear modification on the parton distribution functions.}
\label{fig:ndpf}
\end{figure}

However, experimental data shows that the isolated nucleon picture is too simple \cite{Aubert:1981gv,Bodek:1983qn}.  Quarks and gluons in a nucleon are sensitive to their nuclear environment, so the parton distribution in an isolated proton changes when that proton is inserted into a heavy nucleus, where the quark and gluons from adjacent nucleons are in close proximity.

Figure~\ref{fig:ndpf} shows schematically that there are several different types of modifications to nuclear parton distributions,  occurring at different $x$ values.   At small-$x$ values, interactions reduce the parton densities, a process known as shadowing.   But, due to momentum conservation, if there is nuclear shadowing, there must also be a nuclear enhancement at some other $x$ (\emph{antishadowing}). For the experimentally observed suppression at relatively large $x$ (\emph{EMC effect}) there is as yet no fully accepted theoretical explanation \cite{Thomas:2018kcx}, but short-range nucleon-nucleon correlations are a promising explanation~\cite{Hen:2016kwk}. Finally, the enhancement at $x\sim 1$ is due to the motion of nucleons in the nucleus.

Shadowing occurs in a high-density environments like heavy nuclei, when gluons from nearby nucleons interact.  Two gluons with small $x$ may fuse together, for example, forming a single higher $x$ gluon.  This should happen  most often near the center of the nucleus, and more rarely near its periphery \cite{Emelyanov:1999pkc}.  This makes the gluon distribution of the nucleon to be suppressed at small $x$, compared to free nucleons. 

Traditionally, parton distributions have been measured using deep inelastic scattering (DIS) of high-energy electrons or muons, $l^-A\rightarrow l^-X$, shown in Fig. 2a.  Leptons interact electromagnetically, so DIS only probes the electrically charged quarks.  The gluon densities must be inferred, often by studying how the quarks evolve with the scale $Q^2$ set by the virtuality of the exchanged photon. 
The scale ($Q^2$) dependence of the parton distributions can be calculated by using perturbative QCD (pQCD), but the extraction of the full distributions requires a fit to experimental data. Current fits work at next-to-leading (or even next-to-next-to-leading) order in pQCD. Several such fits exist for protons~\cite{Ball:2017nwa,Hou:2016nqm} and for protons plus nuclei~\cite{Kovarik:2015cma}.    The commonly used EPPS16 fit~\cite{Eskola:2016oht} fits for the nuclear modifications to the CT series fits~\cite{Hou:2016nqm}.  Most of these fits use both DIS data and data on selected hadronic processes at RHIC and the LHC.  The extraction of parton densities from hadronic processes is subject to significant theoretical uncertainties.

Proton structure functions were measured down to $x\approx 10^{-5}$ at the HERA $ep$ collider~\cite{Abramowicz:2015mha}.  But, there is as yet no $eA$ collider, so structure function measurements for ions are limited to lower-energy fixed-target collisions, sensitive mostly to $x>10^{-2}$.  The most energetic fixed-target DIS experiment, Fermilab's E665~\cite{Adams:1995is}, used 470 GeV/c muon to measure shadowing down to $x=3\times 10^{-4}$, but only by going to very low $Q^2$ of 0.1 GeV$^2$, where pQCD is clearly inapplicable.  The HERMES detector at HERA also took data with 27.6 GeV electrons and positrons incident on a series of nuclear gas-jet targets \cite{Airapetian:2007vu}.  They measured both DIS and coherent photoproduction, albeit at energies below some fixed-target experiments. 

Traditional parton distributions give parton distributions in $x$, averaged over the entire nucleus.  One recent study went further, and parameterized parton distributions in terms of impact parameter, based on the $A$ dependence of spatially averaged fits \cite{Helenius:2012wd}.  In order to directly obtain more differential information about the nuclear structure and to directly probe small-$x$ gluons, one needs to go beyond inclusive DIS.

Photoproduction (Fig. 2b) is one such approach.  An incident photon may fluctuate to a quark-antiquark ($q\overline q$) dipole which then interacts with the nucleus.  The simplest dipole-nucleus interaction is single gluon exchange, so the lowest-order cross-section is  directly proportional to the gluon distribution. Single gluon exchange leads to final state such as dijets or mesons containing heavy quarks.  The ATLAS collaboration recently made measurements of dijet production in photon-lead collisions~\cite{ATLAS:2017kwa}.
Single gluons carry color, but the final state particles must be color neutral.  So the reaction requires an additional separate gluonic string between the target and final state.  When this string breaks, it forms mesons and, in the process, generally dissociates the target.  These reactions are experimentally complex, requiring the reconstruction of many-particle final states and accounting for missing momentum due to unseen particles.  
To avoid these complications, most experimental analyses study simpler reactions such as coherent (Fig. 2c) or incoherent (Fig. 2d) photoproduction.  In the coherent case, the final state is exclusive.    Even in incoherent photoproduction, the nuclear breakup often leads to relatively simple final states in which the produced vector meson is clearly separated from the nuclear remnants.
 
Because two gluons are exchanged in these processes, the cross-section is proportional to the square of the gluon density, modulo some theoretical complications.   Photoproduction can be either coherent or incoherent; coherent photoproduction is sensitive to the average gluon distributions, while incoherent photoproduction is sensitive to event-by-event fluctuations in gluon densities. 

Modern studies have gone beyond one-dimension ({\it i. e.} a function of Bjorken$-x$) parton distributions, to consider the spatial dependence of partons within protons -  Generalized Parton Distributions (GPDs) - and Transverse Momentum Distributions (TMDs), which characterize the transverse momentum of partons within a proton.  The ultimate goal, the Wigner function, incorporates information on both the spatial dependence of partons and their transverse momentum; we will discuss one new technique to probe Wigner distributions below.  We will also discuss the nuclear analogs of GPDs - characterizing the spatial distribution of gluons within heavy nuclei - essentially exploring the spatial dependence of shadowing.

In this review, we will discuss some key experimental techniques, before moving on to coherent vector meson photoproduction and the extraction of gluon distributions.  We will then discuss the use of photoproduction to explore the spatial dependence of the parton distributions and event-by-event fluctuations.  We then discuss the very high energy limit, and ways to extract the Wigner distribution, before concluding with a discussion of how parton distributions affect heavy-ion collisions.

\section{Experimental Techniques}

The term photoproduction usually applies to real (with $Q^2=0$) or nearly real photons, while electroproduction is used for virtual photons with $Q^2>0$.   These processes are studied in different types of accelerators.   

In electron-ion collisions, an electron can radiate a virtual photon which then interacts with the target.  At DESY's HERA collider, measurements were made of many processes, including deep inelastic scattering, coherent and incoherent photoproduction of a variety of vector mesons and photoproduction of dijets. Additionally, the point-like structure of the electron makes it possible to study electroproduction \cite{Budnev:1974de}.  By reconstructing the outgoing lepton, the photon kinematics can be completely determined.  HERA collided 27.5 GeV electrons or positrons with 920 GeV protons, reaching $\gamma p$ center of mass energies up to about 300 GeV. 

Unfortunately, HERA never accelerated heavier ions, and ion-target data is much more limited.  Jefferson Lab studies fixed-target electron-ion collisions, but even with the recent upgrade to a 12 GeV electron beam, it is mostly sensitive to valence quark effects, with $x>0.1$.  The U. S. \cite{Accardi:2012qut}, China \cite{Chen:2018wyz} and Europe \cite{AbelleiraFernandez:2012cc} are all considering building an electron-ion collider (EIC) which would allow us to study photo- and electroproduction over a broad range of energies, thereby probing the behavior of quarks and gluons in nuclei at low $x$.

Until an EIC is built, most high-energy photoproduction studies use ultra-peripheral ion collisions (UPCs).  UPCs are interactions that occur at large impact parameters (ion-ion separations), so there are no hadronic interactions to overshadow the electromagnetic processes.  The photons come from the electromagnetic fields of the nuclei \cite{Bertulani:2005ru}.  The photon flux scales as the square of the nuclear charge, so high $\gamma A$ luminosities can be reached, even for moderate $AA$ luminosities.   In the target nucleus rest frame, the maximum photon energy is about $2\gamma^2\hbar c/R_A$, where $\gamma$ is the lab-frame Lorentz boost of one beam and $R_A$ is the radius of a nucleus with atomic number $A$.  At the LHC, target-frame photon energies up to several PeV ($10^{15}$ eV) are accessible, depending on nuclear species, corresponding to $\gamma A$ center of mass energies up to about 1 TeV. This allows access to gluons with $x$ down to  a few $10^{-6}$  \cite{Baltz:2007kq}.  The photon energy $k$ can be inferred from the mass ($M$) and rapidity ($y$) of the final state via $k=M_Vc^2/2\exp{(\pm y)}$.  

In UPCs, the photons are nearly real; their virtuality is limited by the size of the nucleus, $Q^2 \lesssim \hbar^2/R_A^2$, so UPCs probe partons at a scale that is determined by the reaction ({\it e.g.} mass of the produced particle).  

For coherent photoproduction it is impossible to tell which nucleus emitted the photon and which was the target.  This leads to a two-fold ambiguity in photon energy, and, at low transverse momentum, to destructive interference between the two amplitudes~\cite{Klein:1999gv,Abelev:2008ew}.  Proton-nucleus collisions have also been used for UPC studies \cite{Acharya:2018jua}.  These collisions are dominated by photon emission from the nucleus, so mostly probe $\gamma p$ interactions.  Proton-proton collisions have also been studied; the photon flux is lower, but the accessible energies are higher \cite{Klein:2003vd,Aaij:2018arx}.  

UPC photoproduction studies were pioneered by the STAR experiment at RHIC \cite{Adler:2002sc}.  All four large experiments at CERN's Large Hadron Collider (LHC) have also studied UPC final states.  Most analyses have studied simple final states (usually two-prong decays like $\rho\rightarrow\pi^+\pi^-$ or $J/\psi\rightarrow\mu^+\mu^-$), which can be completely reconstructed.  Four-pion states have also been studied, and ATLAS has also probed dijets. 
For coherent photoproduction, the final state transverse momentum satisfies $|\kt| < {\rm few}\ \hbar/R_A \approx 150$ MeV/c.   This distinctive signature allows excellent background rejection; wrong-sign pairs often serve as a measure of background. 

Exclusive production means that there are no other particles in the final state.  In other words, there are rapidity gaps (particle free regions) between both the intact outgoing ions (or the outgoing electron and ion) and the final state.  

In incoherent photoproduction, $p_T$ is higher, and there may be additional particles from the nuclear breakup, so background rejection is harder, particularly for wider resonances.  However, these particles will be in the far forward region, preserving the two rapidity gaps around the vector meson or other final state.

Although these events are distinctive once recorded, triggering is a challenge because of the large backgrounds from beam-gas interactions, low-multiplicity hadronic interactions and even cosmic-ray muons \cite{Nystrand:1998hw}. 

\section{Vector Meson Photoproduction}

In the dipole picture, photoproduction at high energy is seen such that an incident photon fluctuates to a quark-antiquark ($q\overline q$) dipole which then interacts with the nucleus.  The dipole lifetime, $\hbar/M_d$ ($M_d$ is the dipole mass), is generally much longer than the time the dipole spends in the nucleus, so this dipole configuration does not change during the interaction.

Vector mesons have the same quantum numbers as photons, $J^{PC}=1^{--}$, so the photon-nucleus interaction involves elastic scattering \cite{Bauer:1977iq}.  This leads to a large cross-section that, unlike most other high-energy reactions, increases with increasing energy.  In quantum mechanical language, this scattering involves the imaginary part of the nuclear potential. In particle physics language, this can be described as a Pomeron exchange.  In  pQCD, the Pomeron is an infinite gluon ladder, as is shown in. Figure 2d. At lowest order, Pomeron exchange is two gluon exchange. Two gluons is the minimum color-charge neutral exchange.  Photoproduction probes the nuclear structure at a scale set by the mass of the final state.  For vector mesons, $Q^2=(M_Vc^2/2)^2$.  The Bjorken$-x$ value depends on the photon energy, with $x=(M_Vc^2)^2/W^2$, where $W$ is the $\gamma$-nucleon center of mass energy.  
The cross-section for photoproduction of a vector meson $V$ on a nucleus with atomic number $A$ can be written as
\begin{equation}
\sigma(\gamma A\rightarrow VA) = \big|\Sigma_i A_i e^{i \kt \cdot \rt_i/\hbar}\big|^2
\label{eq:coherence}
\end{equation}
where the index $i$ runs over the target nucleons at position $\rt_i$.  The $A_i$ are the production amplitudes on those nucleons (typically they are all the same) and $\kt$ is the momentum transfer from the  nucleus to the vector meson.  When $|\kt| < \hbar/R_A$,  $\exp(i\kt\cdot\rt_i/\hbar)\approx 1$, so $\sigma\propto A^2$ and production is coherent.  At large momentum transfers, the phase factors are large and average out, so $\sigma\propto A$; production is incoherent.

The assumption that the production amplitudes are all the same fails  under closer examination, for two reasons.  First, a single incident $q\overline q$ dipole may interact more than once (but will of course only produce a single vector meson) creating a kind of nuclear (not gluon) shadowing.  This is most likely for photons that pass through the thicker, more central parts of the nucleus.  Second, photoproduction occurs through gluon exchange, and gluon shadowing can reduce the density of gluons.  This shadowing should depend on the nuclear density, and be highest in the central part of the nucleus, and lower in the periphery.   So, depending on their gluon density, not all nucleons will contribute equally to the cross-section.

The degree of shadowing depends on the dipole size: large dipoles have large interaction cross-sections so are heavily shadowed, and will saturate the cross-section (as was seen with proton beams in $pp$ elastic scattering), while small dipoles will see much less shadowing.  Dipole size is inversely related to the photon $Q^2$ and the final state mass, so, it should be possible to observe shadowing at different length scales, including, for small dipoles, gluon shadowing.

\begin{figure}[t]
\begin{center}
\includegraphics[width=0.45\textwidth]{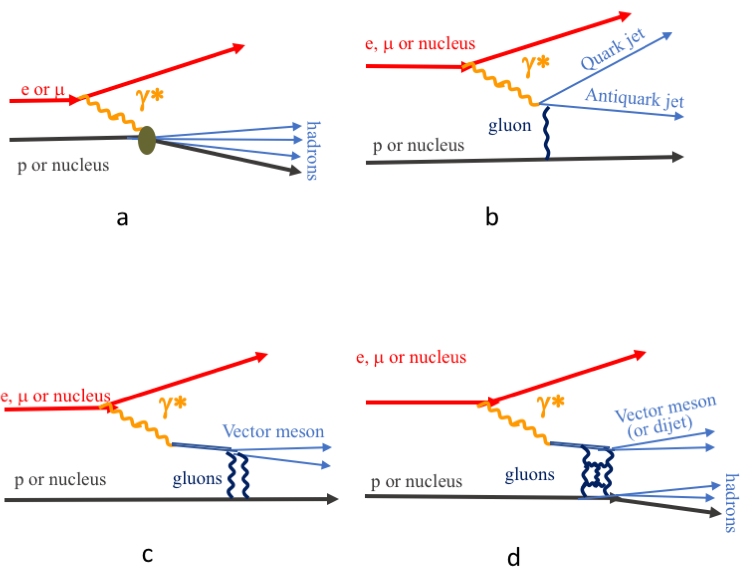}
\end{center}
\caption{Diagrams showing (a) Deep Inelastic Scattering, (b) photoproduction of dijets, (c) coherent vector meson photoproduction and (d) incoherent photoproduction. Panel (c) shows lowest-order 2-gluon exchange, while panel (d) shows a higher-order depiction of a Pomeron as a gluon ladder.}
\label{fig:diagrams}
\end{figure}

In addition to probing the average gluon distributions, photoproduction can also probe event-by-event fluctuation in the target configuration.   Quantum mechanically, photoproduction amplitudes for the different targets only add (Eq. \ref{eq:coherence}) if the nucleus remains in the ground state, {\it i. e.}, if the initial and final nuclear states are the same.  If the nucleus is excited, then the cross-sections are added instead; this is incoherent photoproduction. The Walker-Good approach \cite{Good:1960ba} relates these two cases to the average nuclear configuration (position of nucleons, or quarks and gluons) and to event-by-event fluctuations in those configurations. In this picture, one finds the eigenstates of diffraction, which at high energy are virtual photon fluctuations into a dipole of a fixed size with a given  configuration $\Omega$ for the target nucleus.

In the Walker-Good picture, the total photoproduction cross section is obtained by summing the cross sections for the different eigenstates:
\begin{equation}
\frac{\mathrm{d} \sigma_{\rm tot}}{\mathrm{d}t} = \frac{1}{16\pi} \left \langle \big|A(K,\Omega)\big|^2\right \rangle.
\label{eq:tot}
\end{equation}
Here $A(K,\Omega)$ is the amplitude for the diffractive scattering with the target nucleus in configuration $\Omega$, $t$ is the squared four momentum transfer to the target and $K$ accounts for the kinematic factors in the reaction.  The amplitudes are squared, and then an average $\langle \cdot \rangle$ is performed over target configurations. 
In case of coherent photoproduction, in this picture one sums scattering amplitudes for the different eigenstates and then squares, and obtains
\begin{equation}
\frac{\der\sigma_{\rm coh}}{\der t} = \frac{1}{16\pi} \left|\left\langle A(K,\Omega) \right \rangle \right|^2.
\label{eq:co}
\end{equation}
Averaging over configurations before squaring forces  the initial and final configuratiosn to be the same. The incoherent cross-section is then the difference between Eqs. \eqref{eq:tot} and \eqref{eq:co}:
\begin{equation}
\frac{\mathrm{d}\sigma_{\rm inc}}{\mathrm{d}t} = \frac{1}{16\pi} \bigg(\left\langle \big|A(K,\Omega)\big|^2 \right\rangle - \big|\left\langle A(K,\Omega)\right\rangle \big|^2\bigg).
\label{eq:inc}
\end{equation}
Because of the switched ordering of the squaring and averaging, the incoherent process is sensitive to fluctuations in the nuclear configuration. A more formal treatment of this is given in Refs.~\cite{Miettinen:1978jb,Caldwell:2010zza}.
One unique use of vector meson photoproduction is to study the spin-dependent generalized parton distribution E (GPD-E) by observing $J/\psi$ photoproduction on polarized protons \cite{Lansberg:2018fsy} at RHIC. 

\section{Coherent scattering}

Coherent scattering in the dipole picture is illustrated in Fig.~~\ref{fig:diagrams}c. Two gluons is the minimum color-neutral exchange, required for an exclusive process.

In the dipole picture, the scattering amplitude for the diffractive vector meson production can be written in terms of the dipole-target scattering amplitude $N_\Omega$, which is sensitive to the gluonic density of the target and, in case of proton targets, is constrained by proton structure function measurements~\cite{Rezaeian:2012ji}. It describes how a quark-antiquark dipole interacts with the target, taking into account potential multiple gluon exchanges. The resulting amplitude approximately reads~\cite{Kowalski:2006hc}
\begin{multline}
\label{eq:diffractive_amplitude_dipole}
    A(K,\Omega) =2 \mathrm{i} \int \der^2 \rt  \frac{\der z}{4\pi} \der^2 \bt e^{-i\bt \cdot \kt/\hbar} \\
    \times \Psi^*(\rt,z,Q^2) \Psi_V(\rt,z,Q^2) N_\Omega(\rt,\bt),
\end{multline}
where $\kt$ again is the transverse momentum transfer, and at high energy $|\kt| \approx \sqrt{-t}$. The photon wave function $\Psi^*$ describes the photon $\to$ dipole splitting with the dipole having a transverse size $\rt$ at impact parameter $\bt$ and quark carrying a fraction $z$ of the photon longitudinal momentum, and can be computed using Quantum Electrodynamics. After the scattering process, the dipole forms a vector meson $V$, and this non-perturbative process is described in terms of the vector meson wave function $\Psi_V$. Even though $\Psi_V$ can not be directly calculated, it is constrained by e.g. decay width measurements.

As shown in Eq.~\eqref{eq:co}, one has to calculate the scattering amplitude $\langle A(K,\Omega)\rangle$ averaged over all target configurations $\Omega$. Because the the transverse momentum transfer is Fourier conjugate to the impact parameter, the coherent cross section  measured as a function of momentum transfer probes the transverse density profile of the target proton or nucleus on its average configuration. This makes it possible to do \emph{nuclear imaging} as we discuss in Sec.~\ref{sec:imaging}

The sensitivity to the gluon density is also visible in Eq.~\eqref{eq:diffractive_amplitude_dipole}. The total inclusive cross section can be obtained, by using the optical theorem, from the $\gamma ^*+ A \to \gamma^* + A$ forward elastic scattering amplitude. This amplitude is given by Eq.~\eqref{eq:diffractive_amplitude_dipole}
when the vector meson is a virtual photon ($V=\gamma^*$) and $\kt=0$. Thus, the total cross section is proportional to the dipole amplitude  $\langle A(K,\Omega)\rangle$ (which is proportional to the gluon density), instead of the squared amplitude as in case of an exclusive scattering.

The sensitivity to the gluonic structure is even more transparent in the limit where multiple scattering can be neglected and one describes the vector meson non-relativistically. In these limits, one can recover the Ryskin result for the coherent cross section at small $|t| \ll M_V^2$ at lowest order in pQCD~\cite{Ryskin:1992ui}
\begin{equation}
\label{eq:ryskin}
\frac{\der \sigma_{\rm coh}}{\der t} = \frac{\pi^3 \alpha_s^2 M^3 \Gamma_{V \to e^+e^-}}{3 \alpha_\text{em}} \left[ \frac{1}{(2\bar q^2)^2}  \bar xg(\bar x, \bar q^2)\right]^2 F_N^{2g}(t)^2,
\end{equation}
where $\Gamma_{V \to e^+e^-}$ is the leptonic decay width and the hard scale $\bar q^2 = m_V^2c^4/4$. The cross section is again seen to be proportional to the squared gluon distribution. The phenomenological two-gluon form factor $F_N^{2g}(t)$ describes the distribution of gluons in the nucleus and its Fourier transform is related to the spatial distribution of partons. $F_N^{2g}(t)$ therefore encapsulates the $t$ dependence of the cross-section,

This approach applies equally for single nucleons and for heavier nuclei, as long as one accounts for the total gluon density in the target. Because of the larger size of heavy nuclei, the form factor $F_N^{2g}(t)$ drops much faster with increasing $t$. 

\begin{figure}[t]
\begin{center}
\includegraphics[width=0.45\textwidth]{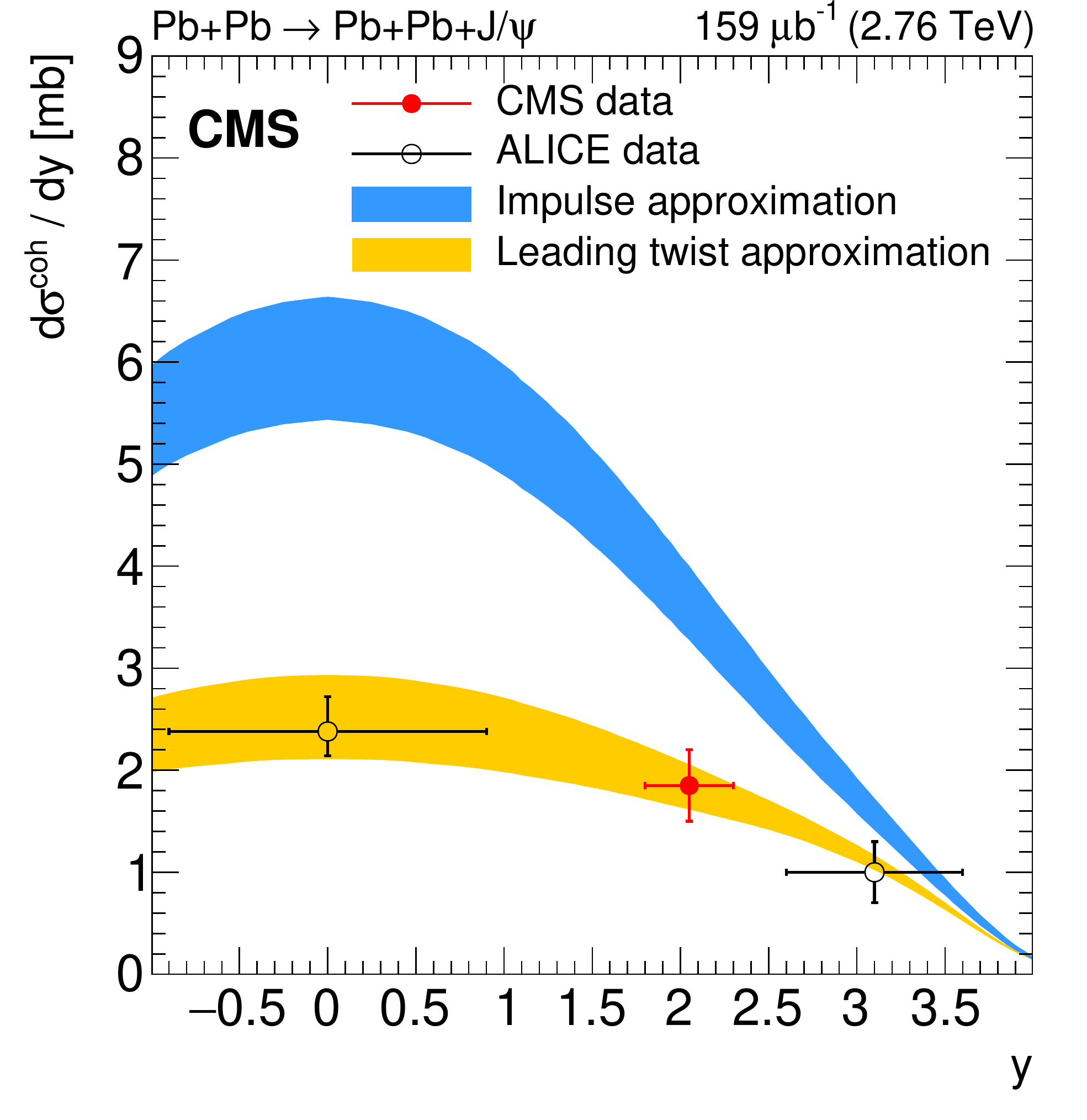}
\end{center}
\caption{Total $J/\psi$ production cross section in UPC Pb+Pb collisions as a function of rapidity, as measured by the CMS collaboration, compared to the impulse approximation prediction (see text). Figure from Ref.~\cite{Khachatryan:2016qhq}.}
\label{fig:impulse_approx_comparison}
\end{figure}

At the LHC, coherent $J/\psi$ photoproduction in lead-lead collisions has been measured by ALICE and CMS experiments~\cite{Abbas:2013oua,Khachatryan:2016qhq}. The simplest way to quantify the magnitude of nuclear effects is to compare the observed $\gamma + \mathrm{Pb} \to J/\psi + \mathrm{Pb}$ cross section to the corresponding cross section off proton targets measured at HERA scaled to the photon-nucleus case by taking into account the nuclear geometry and assuming that there are no additional nuclear effects. This estimate is compared with the CMS measurement~\cite{Khachatryan:2016qhq} in Fig.~\ref{fig:impulse_approx_comparison}.  The suppression is a factor of $\approx 1/3$ compared to the impulse approximation.  The impulse approximation neglects shadowing and other nuclear effects, but does include coherence, so the forward scattering cross-section scales as $A^2$.  The experimental suppression is a clear sign of nuclear effects in the kinematic region probed here: $x \sim 10^{-3}$ and $Q^2 \sim 2.25$ GeV$^2$.

This suppression can be related to the nuclear shadowing. As discussed in Ref.~\cite{Guzey:2013qza}, the observed suppression is compatible with expectations based on nuclear parton distribution fits like EPPS16~\cite{Eskola:2016oht}, assuming that the coherent cross section scales like gluon distribution squared as in Eq.~\eqref{eq:ryskin}.  Unfortunately, the EPPS16 uncertainties for small $x$ gluons with are large, so this consistency is unsurprising.  This shows the benefit that would come from including photoproduction data in nuclear parton fits. 

Similarly, the leading twist nuclear shadowing model~\cite{Frankfurt:2011cs} calculations that are based on QCD factorization theorem and diffractive nucleon parton distribution functions measured at HERA and shown in Fig.~\ref{fig:impulse_approx_comparison} predict the measured nuclear effect on $J/\psi$ production. 

This precise data can provide very powerful constraints on nuclear parton distribution functions. The suppression seen in Fig.~\ref{fig:impulse_approx_comparison} is much larger than the nuclear effects seen in fixed target nuclear DIS like E665~\cite{Adams:1995is}.  The suppression is also larger than is seen in inclusive particle production in proton/deuteron-nucleus collision~\cite{Adler:2006wg}.   In addition, these collisions are more complex because the incident proton has a complicated substructure.  

However, nuclear PDF collaborations do not yet include exclusive vector meson production cross sections in their fits, for a couple of reasons \cite{Klein:2017vua}.  Parton distribution functions are by definition inclusive, and describe interactions with single partons. Exclusive process must be color neutral, so require multiple parton exchange -  two gluons at lowest order.  Fortunately, the dominant kinematic contribution comes when one of the gluons is very soft, so, to lowest order, it can be neglected, or treated with a small correction.  For more accuracy one can treat the two-gluon exchange like a GPD, and make a Shuvaev transformation to treat the gluon $x$ values exactly \cite{Jones:2013pga}.   There are also small corrections to account for photon fluctuations to more complicated states, like $q\overline qg$. Finally, there are uncertainties about the choice of mass scale $\mu$ and also due to the $q\overline q$ dipole to vector meson transition.

Although there is not yet a completely consistent next-to-leading order (NLO) calculation, several recent calculations have taken significant steps in that direction~\cite{Boussarie:2016bkq,Jones:2016ldq}.  The NLO calculations are also sensitive to the quark distributions.  Ref.
\cite{Jones:2016ldq} finds that the major uncertainty is from the mass scale is in the $\pm15\%$ to $\pm 25\%$ range for $\Upsilon$ photoproduction.  Since there is little other data that probes gluon distributions at such low $x$ values, it is important to complete the NLO calculations and also work to reduce the other model uncertainties.

\section{Nuclear Imaging}
\label{sec:imaging}
Experimentally, we can image the nucleus by measuring $\mathrm{d}\sigma/\mathrm{d}t$.  In simple terms, the Fourier transform of $\mathrm{d}\sigma/\mathrm{d}t$ gives the distribution of interaction sites within the nucleus.  The idea of using Fourier transforms to determine transverse nuclear profiles was first applied to proton-proton elastic scattering data, finding an apparent central constant opacity, with a mean interaction radius that increased with collision energy \cite{Amaldi:1979kd}.  
The proton-proton analysis was limited by the inability to control or select the configuration of the incident protons. Photons are simpler initial states, mostly interacting as $q\overline q$ dipoles, and one can choose different dipole size distributions by selecting events based on the final state invariant mass. 

Fourier transforms were first applied to $\gamma p$ collisions to determine proton generalized parton distributions \cite{Diehl}.   At an EIC, one can directly determine $t$, and sharp diffractive minima should be visible, as are visible in Fig.~ 8~\cite{Toll:2012mb}.  In UPCs, $k_z$ is subject to the bi-directional photon directional ambiguity, and the vector meson $\kt$ includes a contribution from the photon transverse momentum.   Generally, $k_z$ is small, so can be neglected, but the photon transverse momentum largely fills in the diffractive minima.  Since the direction of the incident photon  breaks the spherical symmetry of the target, we work in two dimensions.  Then, (see also Eq.~\eqref{eq:diffractive_amplitude_dipole})
\begin{equation}
F(b)\propto \frac{1}{2\pi}
\int_0^{\sqrt{-t_{\rm max}}} \mathrm{d} |\kt| |\kt| J_0(b |\kt|/\hbar)\sqrt{\frac{\mathrm{d}\sigma_{\mathrm{coh}}}{\mathrm{d}t}}.
\end{equation}
where $F(b)$ is the two dimensional profile of the interaction sites, $J_0$ is a Bessel function and $-t_{\rm max}$ the maximum $t$ used.  
There are two important caveats.  First, $d\sigma_{\rm coh}/dt$ exhibits diffractive minima. The sign of the coherent production amplitude changes when crossing these minima.  To account for these changes, the sign of $\sqrt{\mathrm{d}\sigma_\mathrm{coh}/\mathrm{d}t}$ should be flipped when crossing them.  Second, this relationship is exact for $t_{\rm max}=\infty$, but experimental data inevitably has a finite reach in $t$.  Cutting off the transform at an experimentally determined $t_{\rm max}$ will introduce artifacts in the transform.  Understanding and minimizing these artifacts is a significant open challenge.

\begin{figure}[t]
\begin{center}
\includegraphics[width=0.5\textwidth]{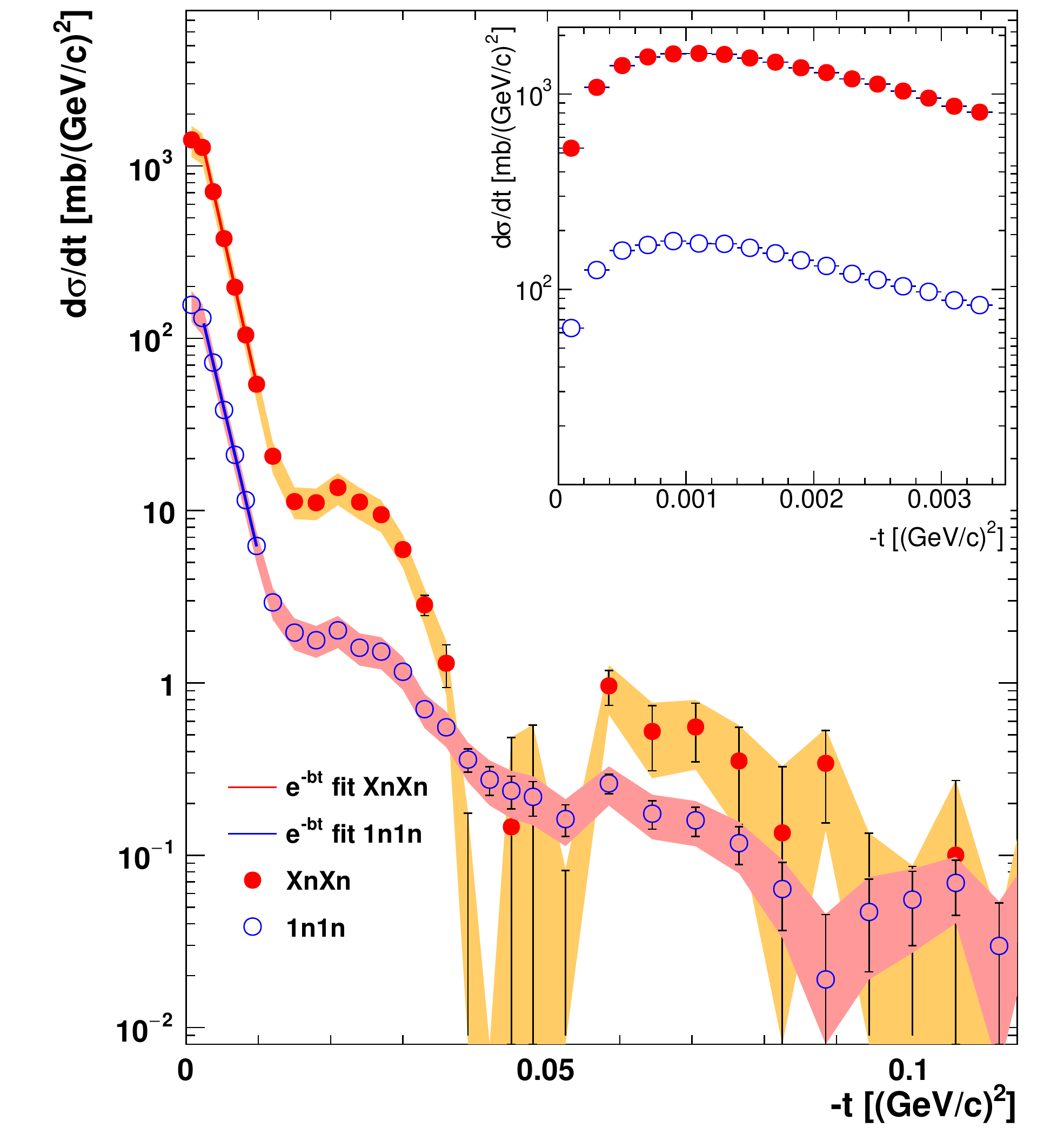}
\end{center}
\caption{Coherent photoproduction $\mathrm{d}\sigma_{\rm coh}
/\mathrm{d}t$ for $\pi^+\pi^-$ photoproduction in 200 GeV Au-Au UPCs, as measured by the STAR Collaboration.  From  Ref. \cite{Adamczyk:2017vfu}.}
\label{fig:dsdtc}
\end{figure}

\begin{figure}[t]
\begin{center}
\includegraphics[width=0.5\textwidth]{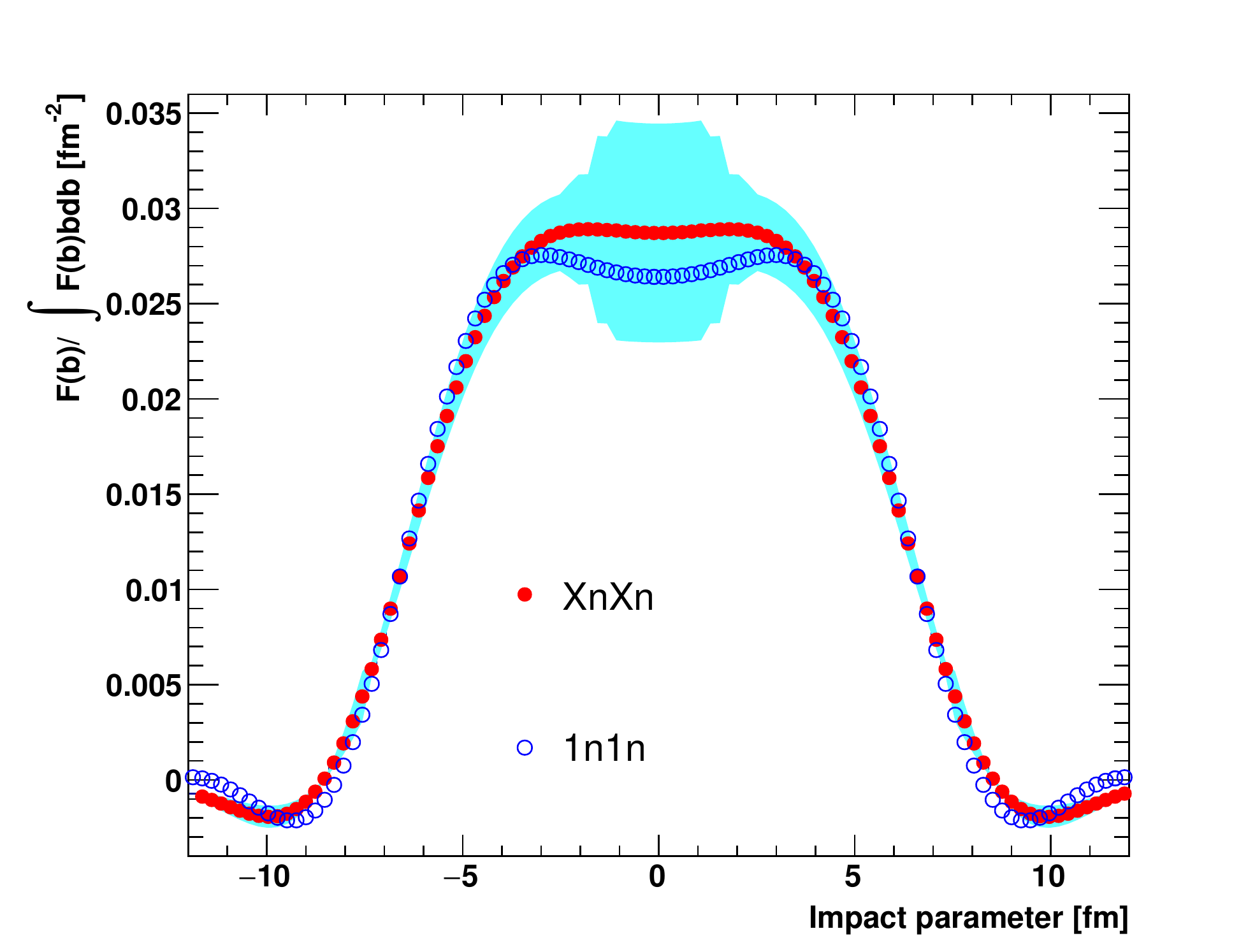}
\end{center}
\caption{Transverse distribution of interaction sites for coherent $\pi^+\pi^-$, as determined by Fourier transformation of Fig. \ref{fig:dsdtc}.  From  Ref. \cite{Adamczyk:2017vfu}.}
\label{fig:fofb}
\end{figure}

The STAR Collaboration made a first measurement of $F(b)$, using 294,000 photoproduced $\pi^+\pi^-$.  They determine the coherent cross-section by first subtracting the background (like-sign pairs) and then subtracting the incoherent photoproduction, as determined from a a dipole form factor fit to $\mathrm{d}\sigma/dt$, made at higher $p_T$, where coherent production is negligible.  The resulting $\mathrm{d}\sigma_{\rm coh}/\mathrm{d}t$ is shown in Fig. \ref{fig:dsdtc}.  They then Fourier transformed the low $-t$ region to find the $F(b)$ in Fig. \ref{fig:fofb}. The shaded band shows the change in $F(b)$ as $-t_{\rm max}$ is varied.  The variation is large at small $|b|$, but the edges of the nuclei are insensitive to the details of the transform.  $F(b)$ is negative at large $|b|$; this is due to a contribution from the other nucleus going in the other direction, which comes in with a negative sign \cite{Klein:1999gv,Abelev:2008ew}.  The dipions were produced by the decay of a state with $J^{PC}=1^{--}$, and the relative rapidity of the two pions was determined by the decay angular distribution \cite{Abelev:2007nb}. 

Recently, STAR has made a preliminary $Q^2$-dependent measurement \cite{Klein:2018grn}.  In it, dipions were divided into three different mass bins, with the dipion mass determining the reaction $Q^2$. 
For a given $-t_{\rm max}$, the shape of $F(b)$ evolved with $Q^2$; for the lightest pairs $F(b)$ was broader, with a flatter top, as expected from nuclear saturation.  However, the variations were comparable to the size of the variations due to changes in $-t_{\rm max}$, so only qualitative comparisons were possible at this point.

\section{Fluctuating parton densities}
  
The parton distribution functions and their multi-dimensional generalizations (including dependence on the parton transverse momentum and/or position) describe the average properties of the hadrons. These averages are subject to quantum fluctuations that make protons look different for each collision. In a nucleus,  fluctuations take place at different distance scales: 1) the positions of the nucleons fluctuate, and 2) each nucleon can have fluctuating partonic substructure.  
At high energies the target configuration is frozen on the time scales of the interaction because of time dilation.

These fluctuations manifest themselves in different observables.  One direct measure of the fluctuations is seen in incoherent photoproduction process in which the target dissociates. As shown in Eq.~\eqref{eq:inc}, the cross section for the incoherent vector meson production depends on the variance of the scattering amplitude, which itself measures the size of fluctuations at distance scale $\sim 1/\sqrt{-t}$ in the target density profile.  Since it takes some energy to excite a nucleus or a proton, in the limit $t\rightarrow 0$, incoherent interactions must disappear.  This corresponds to distance scales much larger than the target, where no internal structure is visible.

Experimentally, these processes are characterized by the presence of a rapidity gap (empty detector) between the dissociated target and the produced vector meson. In UPCs, at small $-t$, coherent scattering dominates over incoherent, and so it may be difficult to separate out small $-t$ incoherent interactions.  Nuclear breakup or proton dissociation can produce particles which go in the far forward region, where instrumentation may be sparse.  In Pb-Pb UPCs, the ALICE Collaboration has obtained ($t$ integrated) cross sections for both processes with moderate accuracy \cite{Abbas:2013oua}.

In electron-ion collisions, detection of the scattered lepton will help ensure that the entire event is seen.  Both coherent and incoherent $J/\psi$ production cross sections in electron-proton collisions were measured relatively accurately at HERA~\cite{Alexa:2013xxa}. However, with the higher luminosities expected at future EICs, event pileup may become problematic. The cross-section to photodissociate a heavy nucleus at an EIC is quite high. Photodissociation can send neutrons in a forward calorimeter, but leave no other signal in the detector, since the electron energy loss is tiny \cite{Klein:2014xoa}.  These neutrons can cause coherent interactions to be mislabelled as incoherent.

\begin{figure}[tb]
\begin{center}
\includegraphics[width=0.5\textwidth]{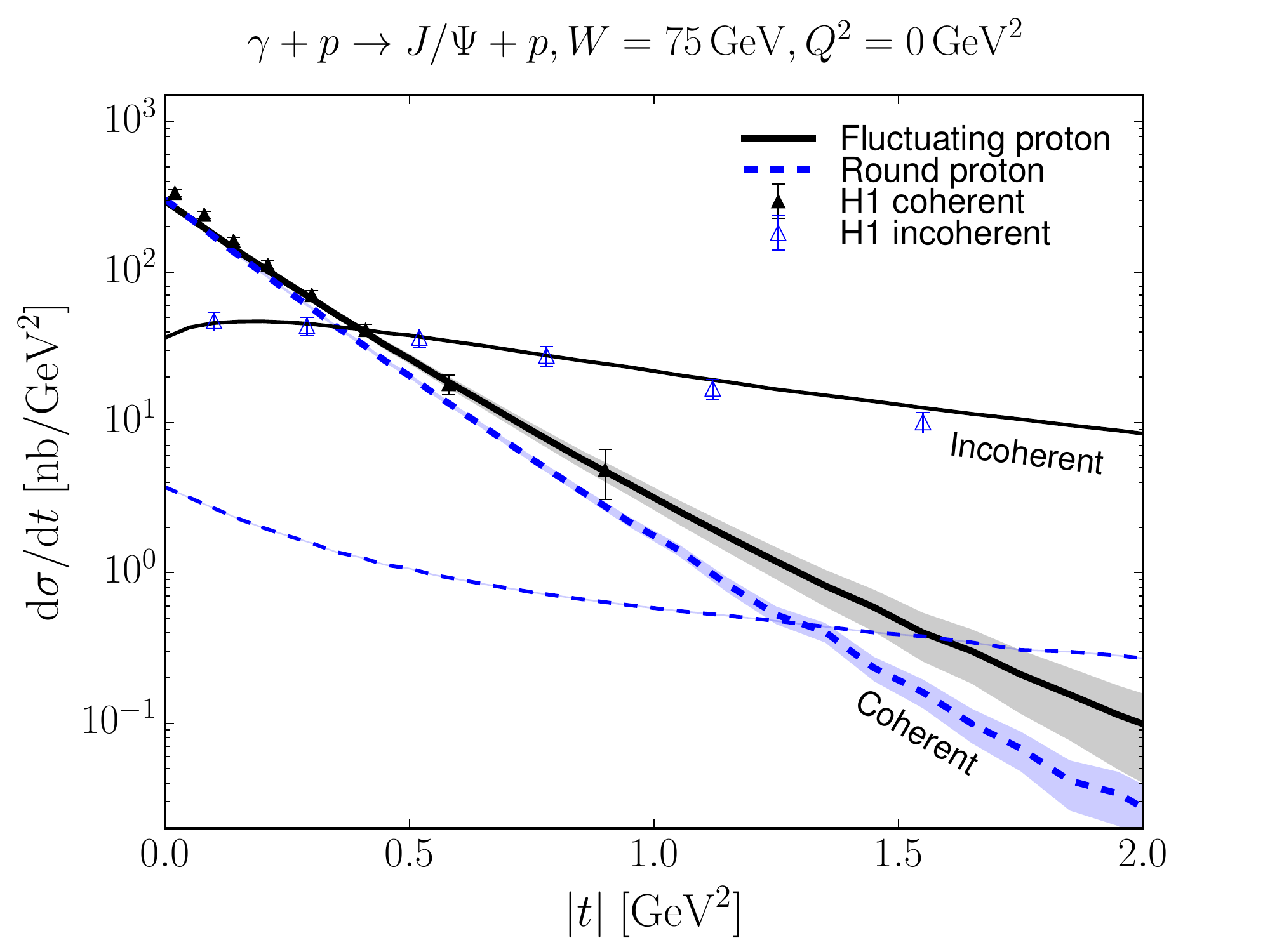}
\end{center}
\caption{Coherent and incoherent $J/\psi$ production cross section as a function of $|t|$ transfer computed using models with and without fluctuations in the proton geometry~\cite{Mantysaari:2016jaz}.}
\label{fig:jpsi_spectra}
\end{figure}

\begin{figure}[t]
\begin{center}
\includegraphics[width=0.5\textwidth]{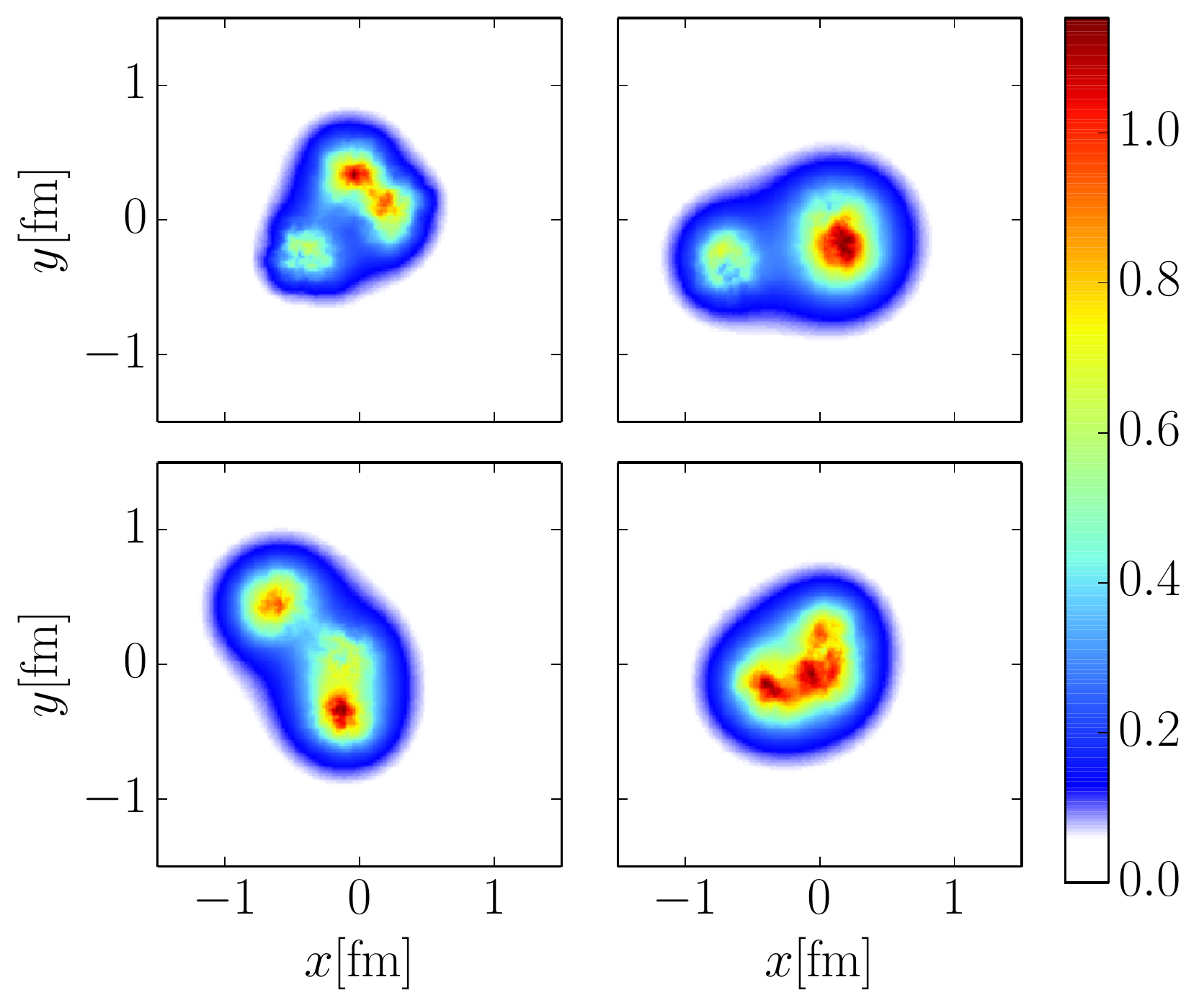}
\end{center}
\caption{Illustration of the extracted proton density fluctuations in arbitrary units on the transverse plane. From Ref.~\cite{Mantysaari:2016jaz}.}
\label{fig:wlines}
\end{figure}

\begin{figure}[t]
\begin{center}
\includegraphics[width=0.5\textwidth]{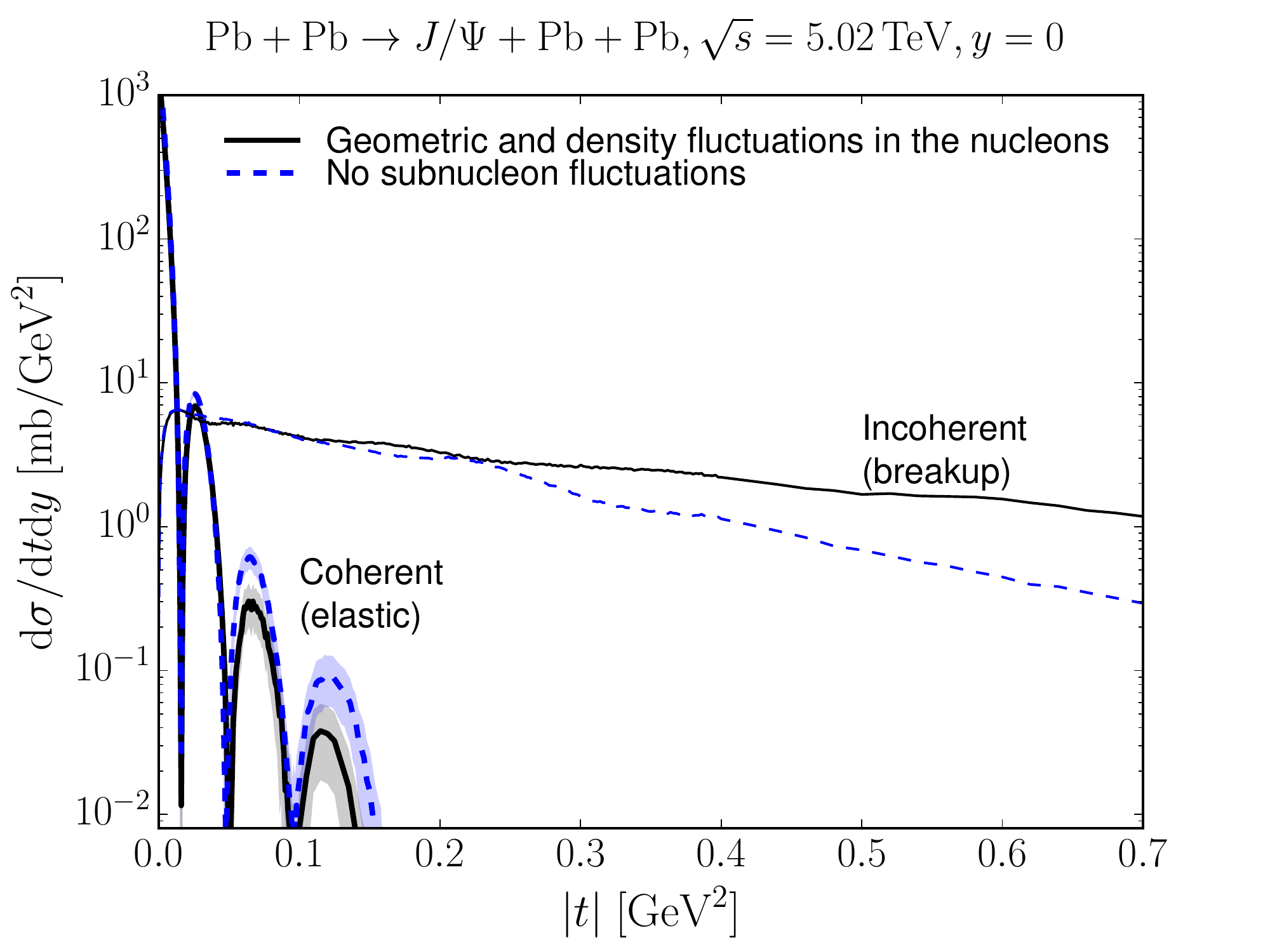}
\end{center}
\caption{Predicted coherent and incoherent $J/\psi$ production cross sections as a function of $|t|$   computed with and without nucleon shape fluctuations From Ref.~\cite{Mantysaari:2017dwh}. }
\label{fig:upc_fluctuations}
\end{figure}

The possibility to use incoherent $J/\psi$ production to probe the proton substructure geometry was first demonstrated in Refs.~\cite{Mantysaari:2016ykx,Mantysaari:2016jaz} where the \emph{ab initio} unknown proton substructure was parametrized as follows: first one assumes that the small-$x$ gluons probed in exclusive vector meson production in HERA kinematics are distributed around three hot spots (e.g. valence quarks). The amount of fluctuations and the size of the proton are controlled by two parameters: the Gaussian width of each hot spot, and the width of the distribution from which the centers for the hot spots are sampled. On top of that, additional overall density fluctuations are implemented.

With this fluctuating proton structure it is possible to calculate both the coherent $J/\psi$ production cross section, which is sensitive to the average shape, and the incoherent cross section, which depends on the fluctuations. Studies of H1 data~\cite{Alexa:2013xxa} found better agreement with models with large event-by-event geometry fluctuations.  Figure~\ref{fig:jpsi_spectra} shows the coherent and incoherent cross sections calculated in the dipole picture using Eq.~\eqref{eq:diffractive_amplitude_dipole}, with the optimal proton shape fluctuations are  compared with the H1 data. For comparison, the same cross sections calculated from the model where there are only color charge fluctuations in the round proton are shown and found to describe only the coherent cross section (the average profile), but to underestimate the incoherent cross section (density fluctuations) significantly. The resulting best-fit proton configurations are illustrated in Fig.~\ref{fig:wlines}.

In nuclei, fluctuations can occur at multiple distance scales. At small $|t|$, $\der\sigma/\der t$ is sensitive to long distance fluctuations, such as  event-by-event fluctuations in nucleon positions from the Woods-Saxon average distribution \cite{Toll:2012mb}. Larger $|t|$ events probe shorter distance scales, such as those from potential nucleon substructure fluctuations.  If nuclei exhibit similar nucleon substructure fluctuations as free protons, then there should be a measurable effect on the incoherent cross section in UPCs at the LHC. This is illustrated in Figure~\ref{fig:upc_fluctuations}, where we show the predicted coherent and incoherent cross sections in ultraperipheral lead-lead collisions ({\it i. e.} photon-nucleus collisions) calculated with and without nucleon shape fluctuations in Ref.~\cite{Mantysaari:2017dwh}. 
The predictions for the incoherent cross section differ at $|t|\gtrsim 0.25\,\mathrm{GeV}^2$, which corresponds to distance scales $\lesssim 0.4$ fm where one is sensitive to nuclear structure at the level of individual nucleons.

Recently there have been improvements over the relatively simple phenomenological model for the fluctuating density profiles presented here. For example, the energy (Bjorken-$x$) dependence of the fluctuating structure of proton and its implications on photo- and electroproduction were studied in Refs.~\cite{Mantysaari:2018zdd,Cepila:2016uku}. The effect of correlations between hot-spot positions was also studied \cite{Traini:2018hxd}. This hot-spot structure is crucial in all theory calculations that describe incoherent vector meson production data from HERA based on a microscopic description of the proton.

\section{Parton densities at high energy}

Parton densities cannot indefinitely continue to increase rapidly with decreasing $x$.  At sufficiently low $x$, the gluon density will reach a maximum, and saturate. Then, a couple of things may happen.

First, at high enough gluon densities, a new state of nuclear matter known as a Color Glass Condensate (CGC) may be formed, where the nucleus can be viewed as a strong classical color field \cite{Gelis:2010nm}. As a result of complex many-body dynamics of the gluon splittings and fusions, an emergent \emph{saturation scale} $Q_s$ is generated. Typical transverse momentum of the gluons in the nuclear wave function becomes of the order of $Q_s$ which increases with decreasing $x$ or increasing energy (which is proposed~\cite{Albacete:2010pg,Lappi:2012nh} to explain the disappearance of the back-to-back correlation in forward dipion production at RHIC~\cite{Adare:2011sc}).  At high enough densities, this scale becomes large compared with the intrinsic QCD scale $\Lambda_\text{QCD}$, which allows perturbative calculations as $\alpha_s(Q_s^2) \ll 1$.  

As gluon densities rise with decreasing $x$, nuclei become increasingly absorptive, and only very small $q\bar q$ dipoles, either from high $Q^2$ photons or the small-dipole extrema from real photons, will be able to penetrate the nucleus.  Small dipoles correspond to high-mass final states, leading to a relative enhancement in high-mass states, such as heavy vector mesons or diffractive dijets, towards small $x$~\cite{Frankfurt:2002wc}.

At sufficiently high densities, the nucleus will begin to look like a completely black disk.  In this limit, event-by-event fluctuations disappear (a black disk is a black disk), so, per Eqs.~\eqref{eq:tot} and \eqref{eq:co}, the total and coherent cross-sections are the same, so incoherent vector meson photoproduction disappears and nuclear interactions become purely absorptive.   The dipole elastic scattering cross-section becomes $2\pi R_A^2$; the '2' here is due to the optical theorem.

Although the overall picture is the same for both proton and nuclear targets, the two species may reach the black disk regime somewhat differently. As we have seen, the gluons in protons appear to contain hot spots.  These hot spots can grow in number and density, and, at sufficiently high energies, encompassing the entire proton.  In contrast, the large number of nucleons  in heavy nuclei may reduce the effects of individual fluctuations at LHC energies, but should not eliminate it.

The energies at which both the CGC and black disk regime appear are not well predicted.  For the black disk regime, the onset energy should increase with increasing vector meson mass ({\it i. e.} decreasing dipole size).  In one calculation \cite{Cepila:2018zky}, the  incoherent (dissociative) $J/\psi$ photoproduction cross section on proton targets reaches a maximum at a photon-proton center of mass energy around 700 GeV, and decreases at higher energies, within the reach of UPC studies at the LHC.   Recent observations of a reduction in neutron production in $J/\psi$ photoproduction at very high energies~\cite{TheALICE:2014dwa} may be a hint of the onset of this regime \cite{Cepila:2017nef}. 
Impact parameter dependent QCD evolution studies also predict a similar experimental signature for proton targets~\cite{Mantysaari:2018zdd}.

\section{Dijets and the Wigner distribution}

\begin{figure}[t]
\begin{center}
\includegraphics[width=0.4\textwidth]{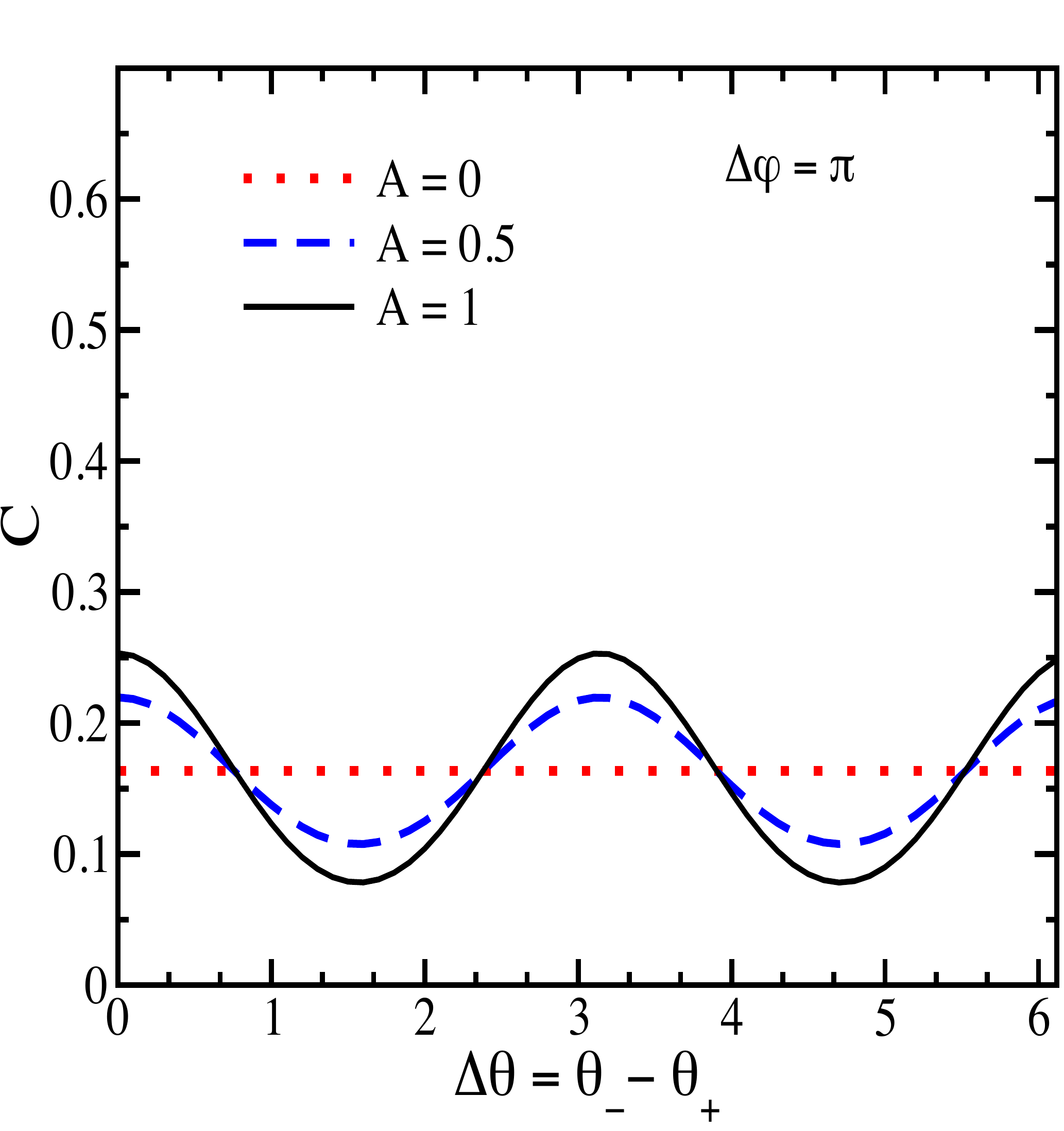}
\end{center}
\caption{Coincidence probability for the production of two jets as a function of azimuthal angle $\Delta \theta$ (see text for details). Figure from Ref.~\cite{Altinoluk:2015dpi}.}
\label{fig:wigner_modulation}
\end{figure}

Going beyond single particle production makes it possible to obtain even more detailed information about the nuclear structure. One powerful process is photoproduction of two jets as show in Fig.~\ref{fig:diagrams}c.  There, the individual jet $p_T$ is much larger than the pair $p_T$.  These two momenta are conjugate to different transverse coordinate vectors.  The very different $p_T$ scales allow for a new type of imaging. 

The ATLAS collaboration has measured inclusive dijet photoproduction (where, in addition to two jets, additional event activity is allowed)~\cite{ATLAS:2017kwa}. Due to the inclusive nature, this process can be related  rigorously to parton distribution functions and can provide useful constraints on nuclear PDF fits at small $x$~\cite{Guzey:2018dlm}. Additionally, taking the advantage of the correlations of different jet momentum variables, it becomes possible to probe the polarization of small-$x$ gluons in the nucleus without polarizing the beam particles~\cite{Dumitru:2018kuw}.  

Dijet photoproduction can look quite similar to diffractive dijet production, since Pomeron exchange can also leave nuclei intact.  Diffractive dijet production dominates for proton-proton interactions, but the cross-section does not rise rapidly with increasing $A$; photoproduction dominates for heavy ions.   The study of diffractive dijet production was pioneered by the CDF experiment at Fermilab's Tevatron \cite{Affolder:2001zn}.  

In exclusive dijet photoproduction, on the other hand, no other processes can take place.   The special advantage of this process is that it can probe the gluon Wigner distribution~\cite{Hatta:2016dxp}, going  beyond what can be expressed in terms of GPDs or TMDs. The Wigner distribution encodes the most complete information about the partonic structure of nuclei, as it  describes not only the spatial distribution of partons, but also how the transverse momentum is distributed between these constituents. However, due to the uncertainty principle, it can not have a probabilistic interpretation similarly as e.g. the parton distribution functions.

In particular, the transverse position and transverse momentum of the parton should be correlated. These correlations lead to an angular correlation between the target recoil momentum $\pt_+ = \pt_1 + \pt_2$ and the mean jet momentum $\pt_- = \frac{1}{2}(\pt_1 - \pt_2)$, with $|\pt_-| \gg |\pt_+|$ (the two jets are almost back-to-back).
If the dijet cross section is measured as a function of the angle $\theta$ between $\pt_+$ and $\pt_-$, this correlation can be quantified in terms of the azimuthal Fourier coefficients $v_n$ when the dijet photoproduction cross section is written as
\begin{equation}
\label{eq:v2}
\frac{\der \sigma^{\gamma^* + A \to j_1 + j_2 + A}}{\der \theta} = v_0[1 + 2v_2 \cos(2\Delta \theta)].
\end{equation}
This decomposition is analogous to the extraction of $v_n$ coefficient in heavy ion physics when studying the properties of the Quark Gluon Plasma (see Sec.~\ref{sec:otherfields}). 

If there is no correlation between the transverse coordinate and momentum, the second Fourier coefficient $v_2=0$, while any correlation gives nonzero contribution to $v_2$~\cite{Hatta:2016dxp}. When the process is viewed in coordinate space, this correlation is a result of having a similar $v_2$ modulation in the dipole-proton scattering amplitude when studied as a function of the orientation of the dipole with respect to the impact parameter: dipoles aligned along the impact parameter scatter with larger probability.

The expected modulation of the cross section is demonstrated in Fig.~\ref{fig:wigner_modulation}, where the coincidence probability for the exclusive two jet photoproduction cross section is shown as a function of $\Delta \theta$ as predicted in Ref.~\cite{Altinoluk:2015dpi} (see also Ref.~\cite{Mantysaari:2019csc} for a more detailed analysis in the same framework). In that analysis,
the dependence of the dipole-target scattering amplitude on the angle between dipole size $\rt$ and impact parameter $\bt$ parametrized by the coefficient $A$, with $A=0$ corresponding to the case where there is no dependence and non-zero $A$ refers to stronger angular modulation.
In momentum space, nonzero $A$ corresponds to a non-trivial angular correlation between the gluon transverse coordinate and its transverse momentum in the wave function of the proton.

The exclusive dijet production has not yet been measured in any collider experiment. As it is currently the only process which is shown to be directly sensitive to the gluon Wigner distribution, there is strong motivation to attempt to measure it in the future DIS experiments at the Electron Ion Collider or at the LHeC, also with nuclear targets.  The ATLAS collaboration has measured inclusive dijet production in UPCs at the LHC~\cite{ATLAS:2017kwa}. It remains to be seen if a similar measurement is possible with exclusive dijets.

\section{Impact on other fields}
\label{sec:otherfields}
Nuclear parton densities are important for two related fields: heavy-ion physics, and cosmic-ray physics.  Parton distributions determine the initial state of relativistic heavy-ion collisions. The small$-x$ behavior affects soft particle production in the collisions, while the spatial dependence of the parton distributions affects the centrality dependence of key observables \cite{Emelyanov:1999pkc}.  Parton distributions are also important to understanding cosmic-ray air showers, where the assumed behavior of partons with low momentum fractions affects the high-energy cosmic-ray composition inferred from data.

\subsection*{Heavy ion physics}

Understanding partonic content of nuclei is necessary for proper interpretation of heavy ion collisions. When two nuclei collide at high energy, they form a hot and dense blob of QCD matter, known as the Quark-Gluon Plasma (QGP). Understanding the properties of this new state of matter, which also formed immediately after the Big Bang, is a key task for the RHIC and LHC heavy ion programs.

The QGP has been found to be an almost perfect (very low viscosity to entropy density ratio) fluid, and its evolution can be described by applying relativistic hydrodynamics. However, these simulations require as an input the initial state of the evolution, which depends strongly on the distribution of partons in the nucleus on an event-by-event basis. In many hydrodynamical simulations like~\cite{Paatelainen:2012at}, nuclear parton distribution functions are an important ingredient in the initialization of the plasma evolution. Understanding the possible non-trivial nucleon geometry (geometry fluctuations) is also important, as hydrodynamics transforms initial state spatial anisotropies into momentum space correlations. These correlations are measured and quantified as Fourier coefficients $v_n$ of the particle distributions in the azimuthal angle similarly as in Eq.~\eqref{eq:v2}.

The initial state geometry fluctuations are especially important in proton-nucleus collisions, where the initial energy density distribution on the transverse plane is determined dominantly by the proton density profile~\cite{Dusling:2015gta}. If the proton were round not only in average but also in each event, the initial state eccentricity would be small and measured $v_n$ ($n \ge 2)$ coefficients approximately zero. This model is contradicted by LHC measurements (e.g.~\cite{ABELEV:2013wsa,Chatrchyan:2013nka,Aad:2014lta}). On the other hand, if the proton and nucleons in the lead nucleus have geometry fluctuations determined from diffractive $J/\psi$ production as discussed above, a good description of the LHC $v_2$ and $v_3$ measurements is obtained~\cite{Mantysaari:2017cni}. 

Despite the success of the hydrodynamical simulations and the \emph{hot spot picture}, we note that there are other explanations for the azimuthal anisotropies seen in proton-nucleus collisions that do not rely on hydrodynamical evolution. For example, the Color Glass Condensate initial state description can produce the same systematic trends for the $v_n$ harmonic coefficients as observed in experiments~\cite{Mace:2018vwq}. Further, the applicability of hydrodynamics to small systems like those produced in proton-nucleus collisions is not yet completely established~\cite{Niemi:2014wta,Gallmeister:2018mcn}.

\subsection*{Cosmic ray physics}

Knowledge of low-$x$ parton densities is critical for understanding ultra-high energy cosmic ray air showers, where an ultra-high energy nucleus strikes the top of the atmosphere.  Air showers with energies up to $3\times10^{20}$ eV have been observed \cite{Bird:1994mp}.  These showers are observed with air fluorescence detectors, which measure the spatial development of the resulting particle shower, with surface detectors which study the particles reaching the surface, and with buried muon detectors.  Interpreting these data requires a model of hadronic air shower development.  This is critically important for studies of the cosmic-ray composition, where we try to determine, based on these observables, whether the incident nucleus is a proton or heavier nucleus.  The speed with which the shower develops as it propagates through the atmosphere depends on the interaction cross-section, and on the characteristics of these interactions, particularly the inelasticity, the fraction of its incident momentum retained by the incident particle, and on the number of lower energy particles produced in each collision.  Current models fail, in that they tend to predict fewer ground-level muons than are seen in the data  \cite{Aab:2016hkv,Abbasi:2018fkz}.  

The forward particle production that determines what particles in an air shower reach the ground has been studied in high-energy $pp$ and $pPb$ collisions by the LHCf experiment \cite{Tricomi:2018fek}.  The collaboration found that most string and Pomeron based Monte Carlo models do not to do a good job of reproducing their data, at least partly due to inadequate treatment of diffractive events.  Although this data does provide some significant constraints, there are many ways to tune the simulations to better match the data, and the LHCf does not point to a unique solution.  It is also difficult to accurately scale in energy and to interpolate between $pp$ and $pPb$ collisions to understand oxygen and nitrogen targets. A more first-principles based approach is desirable.

The most important interactions involve high$-x$ partons from the incident cosmic-ray interacting with low$-x$ partons in nitrogen and oxygen in the atmosphere.    Nuclear phenomena that affect low$-x$ partons at the initial states of shower development could ground based observables, and measurements of composition dependence.  This could affect the altitude at which showers from a given incident nucleus reach their maximum extent \cite{Pajares:2000sn}.  If nuclei look like black disks at sufficiently low $x$, then this would also impact shower development, leading to an increase in interaction inelasticity - the incident particle would retain more energy - and making the shower develop more gradually.  Saturation has been included in some air shower models \cite{Pierog:2006qv}, but currently this amounts to adding additional free parameters- we desperately need data to determine when (in $x$ and $Q^2$) saturation is important, and how it changes with these variables.   

An alternate approach to study composition is to probe a region where pQCD is applicable.   The laterally separated TeV-energy muons seen by IceCube \cite{Abbasi:2012kza} are one example.  These muons have transverse momentum $p_T > 2$ GeV/c, and, because of their high energies, must come the decays of hadrons that are produced early in the shower.  So, these muons should be describable by pQCD calculations.  For these calculations, it is important to understand the possible effect of saturation on the low$-x$ partons in the nitrogen/oxygen targets.

\section{Conclusions}

Modern pictures of protons and heavier nuclei go far beyond the traditional 3 valence quarks + gluons + sea quarks in each nucleon.  We now know that partons within nucleons can interact with nearby partons, leading to density suppression in the core of nuclei, leading to a spatially uneven parton density.  We have also started to make multi-dimensional images of parton densities, studying how the parton distributions depend on transverse spatial position within the nucleus, or with transverse momentum.  We have even started to measure how parton densities fluctuate with time, by measuring event-by-event fluctuations in the distributions.

Looking ahead, over the next decade, the LHC runs 3 and 4 should produce a very large increase in data volumes, enabling high-statistics studies of many UPC topics \cite{Citron:2018lsq}.  High statistics measurements of $J/\psi$, $\psi'$ and $\Upsilon$ states, coupled with further progress in NLO calculations should lead to precise shadowing measurements which should be included in nuclear parton distribution fits.   Data from RHIC on $J/\psi$ photoproduction would help fill in some points at larger $x$.  It may also be possible to measure GPD-E by studying $J/\psi$ photoproduction in collisions involving polarized protons.

This should be supplemented by additional LHC data on dijet photoproduction, and, hopefully photoproduction of open charm, and, possibly, bottom \cite{Klein:2002wm,Strikman:2005yv,Goncalves:2017zdx}.  Charm and bottom hadrons are produced at relatively high rates, and their detection will benefit from the next generation of LHC vertex detectors.  Open charm is of particular interest because it can be probed down to lower masses, so is sensitive to partons at lower $x$ and $Q^2$ than dijets.  Together, these measurements should give us a reasonably clear picture of shadowing at low and moderate $x$ values.  Studies involving an intermediate nucleus will be important in understanding how shadowing evolves with atomic number.  A proposed proton-oxgyen run would be particularly useful for cosmic-ray studies
\cite{Dembinski:2019uta}.  The dijet data should enable the first measurements of the Wigner distribution. 

$\mathrm{d}\sigma/\mathrm{d}t$ based tomography is in its infancy, and will benefit from theoretical developments and from more data. It is particularly important to extend measurements of coherent photoproduction to larger $|t|$.  This would benefit greatly from triggers that do not require neutron emission.

Somewhat further ahead, a future electron-ion collider will provide a huge increase in both the amount and quality of data.  Electron-nucleon scattering allows experimenters to vary the photon $Q^2$ by selecting events based on the recoiling electron, while keeping other important parameters fixed.  So, it could see how the nuclear shape varies with $Q^2$ for a fixed dipion mass, or for $J/\psi$ photoproduction, removing any uncertainty due to changing hadronic physics.   It will also allow us to make a much cleaner determination as to whether the target nucleus remained intact, improving our measurements of incoherent photoproduction.  Finally, it will produce billions of dipions, 10s of millions of $J/\psi$ and of order 100,000 $\Upsilon$~\cite{Lomnitz:2018juf} in a year, allowing for detailed measurements of coherent and incoherent photoproduction over a wide range of $Q^2$.  

Detectors at an EIC will need a very large angular acceptance to detect the scattered electron over the full range of $x$ and $Q^2$.  In interactions at low $Q^2$, the scattered electron is scatters by a very small angle.  The detectors for these electrons must be very close to the beam.  At very high $Q^2$, the electron can be scattered backwards.  Over a broad angular range, good particle identification capability is required to separate the scattered electron from the charged particles produced in the interaction. 

\section*{Acknowledgments}
H.M. is supported by the Academy of Finland, project 314764, and by the European Research Council, Grant ERC-2015-CoG-681707.  
SK's work was funded by the U.S. DOE under contract number DE-AC02-05-CH11231.

\bibliographystyle{apsrev4-1} 
\bibliography{refs}

\end{document}